\documentstyle[sprocl]{article}
\def\be{\begin{equation}}
\def\ee{\end{equation}}
\def\bea{\begin{eqnarray}}
\def\eea{\end{eqnarray}}

\begin{document}
\vbox{
\halign{&#\hfill\cr
        & MADPH-96-959 \cr } }
\bigskip
\title{NRQCD Prediction for the Polarization 
of the $J/\psi$ Produced from b-decay }  
\author{Sean Fleming}
\address{Department of Physics, University of Wisconsin, 
Madison, WI 53706, USA}
\maketitle
\abstracts{Presented at the Meeting of the Division of Particles and
Fields, August 1996, in Minneapolis. The work presented here is based
upon a recent paper\footnote{(hep-ph/9608413)} done in collaboration with Oscar
F. Hern\'andez, Ivan Maksymyk, and H\'el\`ene Nadeau. The NRQCD
predictions for the polarization of the $J/\psi$ produced in $b \to
J/\psi +X$, as well as the helicity-summed production rate are presented.}

A rigorous theoretical framework within which quarkonium production can 
be studied is provided by the NRQCD factorization 
formalism~\cite{bbl}. This 
approach is based on NRQCD, an effective field theory that
that can be made equivalent to full QCD to any desired order in the 
heavy-quark velocity $v$. The NRQCD Lagrangian is 
\bea
{\cal L} & = &
- {1 \over 2} {\rm tr} \, G_{\mu \nu} G^{\mu \nu}
  \;+\; \sum \bar \Psi_\ell \; i {\not \! \! D} \Psi_\ell 
\nonumber \\
& & + \; 
\psi^\dagger \left(i D_0 + {{\bf D}^2 \over 2 m_Q} \right) \psi
\;+\; 
\chi^\dagger \left(i D_0 - {{\bf D}^2 \over 2 m_Q} \right) \chi 
\; + \; \cdots
\label{NRQCDlagrangian}
\eea
where $G_{\mu \nu}$ is the gluon field-strength tensor, $\Psi_\ell$ is
the Dirac spinor field for a light 
quark, $D^\mu = \partial^\mu + i g A^\mu$ is the gauge-covariant
derivative (with $g$ being the QCD coupling constant), $m_Q$ 
is the heavy quark mass, and $\psi$ and $\chi$ are 2-component fields
that annihilate heavy quarks and create heavy antiquarks respectively.  
Color and spin indices on the fields have been suppressed. 
Eq.~\ref{NRQCDlagrangian} describes a fully relativistic field
theory for the light degrees of freedom coupled with a Schr\"odinger field
theory for the heavy quarks and antiquarks. The relativistic effects
of full QCD are reproduced through additional terms represented 
by the $\cdots$ in Eq.~\ref{NRQCDlagrangian}.
In principle there are infinitely many terms. However, using NRQCD $v$-scaling
rules it is possible to retain only those that contribute through a
given order in $v$.  

The NRQCD factorization formalism is a rigorous derivation within
NRQCD of a factored form for quarkonium production and decay rates.
A central result is that inclusive
quarkonium production cross sections must have the form 
\be
\sigma(A+B \to H + X) = \sum_n \frac{F_n}{m_Q^{d_n - 4}}
\langle {\cal O}^H_n \rangle \; ,
\label{NRQCDproduction}
\ee
where the ${\cal O}^H_n$ represent NRQCD four-fermion production
operators, with the index $n$ labeling color and angular-momentum quantum
numbers. The NRQCD matrix elements 
{}$\langle {\cal O}^H_n \rangle \equiv \langle 0| {\cal O}^H_n | 0 \rangle$,
parameterize the hadronization of a heavy-quark-antiquark pair with quantum
numbers $n$ into a quarkonium
state $H$. They scale as the parameter $v$ raised to a power. 
The short-distance coefficients $F_n$ are obtainable
in perturbation theory as a series in $\alpha_s(m_Q)$. 
Information on the order in $v$ of the matrix elements, 
along with the dependence
of the $F_n$ on coupling constants, 
permits us to decide which terms must be retained in
expressions for observables to reach a given level of accuracy. 

Consider the specific case of $J/\psi$ production. In many
instances the most important NRQCD matrix elements are 
{}$\langle{\cal O}^\psi_1({}^3S_1) \rangle$, 
{}$\langle{\cal O}^\psi_8({}^3S_1) \rangle$, 
{}$\langle{\cal O}^\psi_8({}^1S_0) \rangle$, and  
{}$\langle{\cal O}^\psi_8({}^3P_J) \rangle$. 
Inclusive production of $J/\psi$ {}from $b$-decay
provides two measurable combinations of these matrix elements. The first
one is the helicity-summed rate 
$\Gamma(b \to J/\psi + X)$.  
The second combination concerns the polarization parameter
$\alpha$ appearing in the electromagnetic decay rate of $J/\psi$ to
lepton pairs:
\begin{equation}
\label{thetadefined}
\frac{d\Gamma}{d\cos\theta}\big( \psi \rightarrow \mu^+\mu^-(\theta) 
\big) 
\propto 1 + \alpha \cos^2{\theta} \; ,
\end{equation}
where the polar angle $\theta$ is defined in the $J/\psi$ rest frame for
which the $z$-axis is aligned with the direction of motion of the
$J/\psi$ in the lab.

The branching ratio for $b \to J/\psi + X$ is calculated using a matching 
procedure~\cite{makflem}. One obtains~\cite{fhmn}
\begin{eqnarray}
\label{totalfacform}
\lefteqn{BR(b \rightarrow J/\psi + X)  = }
\nonumber \\
& & 0.002      \big\langle{\cal O}^\psi_1({}^3S_1)\big\rangle 
        + 0.1  \big\langle{\cal O}^\psi_8({}^3S_1)\big\rangle
        + 0.2  \big\langle{\cal O}^\psi_8({}^1S_0)\big\rangle 
       +  0.6 {\big\langle{\cal O}^\psi_8({}^3P_0)\big\rangle
                           \over m^2_c}  \; .
\end{eqnarray}
Only the leading color-singlet piece and the leading
color-octet pieces in the relativistic $v^2$-expansion are
considered. The above formula concurs with previous results~\cite{kls}.  

According to the NRQCD $v$-scaling rules, the color-octet matrix
elements in Eq.~\ref{totalfacform}
are all expected to be suppressed by $v^4$ 
with respect to $\langle{\cal O}^\psi_1({}^3S_1) \rangle$.  
However the short-distance coefficients 
in the color-octet terms are some 50 times larger than the
color-singlet coefficient!

Beneke and Rothstein \cite{br} and Braaten and Chen \cite{bc} have
developed techniques for deriving the production rates of 
quarkonia with specified helicities.  Applying their methods one obtains
an expression for the polarization parameter~\cite{fhmn}  
\be
\label{alpha}
\alpha = \frac{
         -0.39 \big\langle{\cal O}^\psi_1({}^3S_1) \big\rangle 
     -   17    \big\langle{\cal O}^\psi_8({}^3S_1) \big\rangle
     +   52    \big\langle{\cal O}^\psi_8({}^3P_0) \big\rangle / m^2_c 
}  
{
              \big\langle{\cal O}^\psi_1({}^3S_1) \big\rangle 
     +  44    \big\langle{\cal O}^\psi_8({}^3S_1) \big\rangle 
     +  61    \big\langle{\cal O}^\psi_8({}^1S_0) \big\rangle 
     + 211    \big\langle{\cal O}^\psi_8({}^3P_0) \big\rangle / m^2_c 
} \;  .
\ee

While experimental determinations of
helicity-summed $BR(b\to J/\psi + X)$ 
have already been carried out~\cite{pdb}, a
measurement of the polarization parameter
$\alpha$ is not yet available. 
Anticipating the availability of this latter measurement, it is
interesting to determine the range of $\alpha$ which is
consistent with existing information on the matrix elements 
{}$\langle{\cal O}^\psi_1({}^3S_1) \rangle$,
{}$\langle{\cal O}^\psi_8({}^3S_1) \rangle$,
{}$\langle{\cal O}^\psi_8({}^1S_0) \rangle$, and 
{}$\langle{\cal O}^\psi_8({}^3P_0) \rangle/m_c^2$,
and with various constraints on linear combinations of these
quantities~\cite{constraints}. The {\it maximum} value for $\alpha$
is $-0.09$, and the {\it minimum}
value for $\alpha$ is $-0.33$.

Unfortunately, due to the large number of poorly determined
parameters the expected range of $\alpha$ as predicted by NRQCD is
large. 
It may be that an accurate NRQCD prediction of the polarization
parameter (beyond leading order) will not be possible for a long
time. This possibility, however, does not degrade the importance of
this calculation, since an experimental determination of $\alpha$ ---
in conjunction with these results --- will most certainly serve to
tighten the constraints on the possible values of the color-octet
matrix elements. In fact, the NRQCD prediction for $\alpha$ is very
sensitive to the values of the color-octet matrix elements. This
offers the hope that a measurement of the polarization of $J/\psi$
produced in $b \to J/\psi+X$ will be instrumental in determining
$\langle{\cal O}^\psi_8({}^3S_1) \rangle$, $\langle{\cal
O}^\psi_8({}^1S_0) \rangle$, 
and $\langle{\cal O}^\psi_8({}^3P_0) \rangle$.

\bigskip

This work was supported by the U.S.~D.O.E.
under Grant No.~DE-FG02-95ER40896, by the University of Wisconsin Research
Committee with funds granted by the Wisconsin Alumni Research
Foundation.  
\section*{References}


\vfill\eject

\end{document}